# A cosmic hall of mirrors

Jean-Pierre Luminet
Laboratoire Univers et Théories (LUTH) – CNRS UMR
Observatoire de Paris, 92195 Meudon (France)
Jean-pierre.luminet@obspm.fr

**Abstract**

Conventional thinking says the universe is infinite. But it could be finite and relatively small, merely giving the illusion of a greater one, like a hall of mirrors. Recent astronomical measurements add support to a finite space with a dodecahedral topology.

**Introduction**

For centuries the size and shape of the universe has intrigued the human race. The Greek philosophers Plato and Aristotle claimed that the universe was finite with a clear boundary. Democritus and Epicurus, on the other hand, thought that we lived in an infinite universe filled with atoms and vacuum. Today, 2500 years later, cosmologists and particle physicists can finally address these fundamental issues with some certainty.

Surprisingly, the latest astronomical data suggest that the correct answer could be a compromise between these two ancient viewpoints: the universe is finite and expanding but it does not have an edge or boundary. In particular, accurate maps of the cosmic microwave background – the radiation left over from the Big Bang -- suggest that we live in a finite universe that is shaped like a football or dodecahedron, and which resembles a video game in certain respects.

In such a scenario, an object that travels away from the Earth in a straight line will eventually return from the other side and will have been rotated by 36 degrees. Space might therefore act like a cosmic hall of mirrors by creating multiple images of faraway light sources, which raises new questions about the physics of the early universe. However, this is just one possibility and other proposals made by researchers in the expanding field of cosmic topology include tetrahedral and octahedral spaces, flat doughnuts and an infinite "horn-shaped" universe.

**The curvature of space**

The first testable predictions about the size and shape of the universe were made by Einstein in 1916 as part of his general theory of relativity. In general relativity massive bodies such as stars change the shape of space--time around them, much as a bowling ball would change the shape of a trampoline. Indeed, it is this local deformation of space--time that is responsible for gravity in Einstein's theory.

The average curvature of space therefore depends on the overall density of matter and energy in the universe. This density is usually expressed in terms of the parameter $\Omega$, which is defined as the ratio of the actual density to the critical density required for space to be flat or Euclidean. Space can therefore have three possible curvatures: zero curvature ($\Omega = 1$), which means that two parallel lines remain a constant distance apart as they do in the familiar Euclidean space; negative curvature ($\Omega < 1$), with parallel lines diverging as they do on the hyperbolic surface of a saddle; or positive curvature (Ω > 1), which means that parallel lines eventually cross one another as they do on the surface of a sphere.

In the standard model of cosmology space has been flat and infinite ever since the universe underwent a short period of extremely rapid expansion called inflation shortly after the Big Bang. Moreover, we now know that the expansion of the universe is actually accelerating due to a mysterious repulsive force caused by "dark" energy (see *Physics World* May 2004 pp37--42).

In 2003 the Wilkinson Microwave Anisotropy Probe (WMAP) produced a high resolution map of the cosmic microwave background that provided clues about the expansion rate of the universe and its curvature. Combined with other astronomical observations the WMAP data suggest that Ω = 1.02 ± 0.02, which favours a spherical universe with positive curvature. The simplest such space is a hypersphere, which can be thought of as the 3D surface of a 4D ball, just as an ordinary sphere is the 2D surface of a 3D ball. Hyperspherical space is therefore finite but it does not have a problematic boundary (figure 1). However, as we will see, many other spherical spaces can fit the data better than a hypersphere.

**The topology of space**

Curvature is clearly central to the large-scale shape of space, but it is not the only factor. The global topological properties of space are also important because they determine whether the universe is finite or infinite. All spherical spaces are finite, for instance, but not all finite spaces are spherical. Indeed, flat and hyperbolic spaces can have finite *or* infinite volumes depending on their topologies.

To illustrate this in two dimensions, think of a square and identify opposite sides as being the same, as happens in video games where a spaceship disappearing to the right of the screen reappears on the left. In three dimensions a spaceship or anything else (such as a particle or a photon) that leaves the "fundamental" cube through one face re-enters it from the opposite face. In this case one can imagine a cubical block of space whose opposite faces have been "glued" together to produce what is effectively a 3D torus.

At first glance all the familiar rules of Euclidean geometry hold in both of these examples, and the spaces looks infinite to those who live in them. However, unless the spaceship keeps encountering the same objects on its travels it has no way of telling if it is moving through an infinite space or through the same finite space again and again.

General relativity does not distinguish between these possibilities because each of the three plausible cosmic geometries -- flat, hyperbolic and spherical -- is consistent with many different topologies. For example, a 3D torus and ordinary flat Euclidean space are described by the same equations in general relativity, even though the former is finite and the latter infinite. Determining the topology of the cosmos therefore requires some physical understanding beyond relativity, in particular concerning the way different regions of space--time are connected.

Cosmologists usually assume that the universe is simply connected like a plane, which means there is only one direct path for light to travel from a source to an observer. A simply-connected Euclidean or hyperbolic universe would indeed be infinite, but if the universe is multiply-connected like a torus there would be many different possible paths. This means that an observer would see multiple images of each galaxy and could easily misinterpret them as distinct galaxies in an endless space, much as a visitor to a mirrored room has the illusion of seeing a crowd. Could we, in fact, be living in such a cosmic hall of mirrors?

Topologists have proved that in addition to the ordinary, simply-connected Euclidean, spherical and hyperbolic spaces there are 17 other Euclidean spaces and an infinite number of spherical and hyperbolic spaces -- all of which are multiply connected. These spaces differ in the shape of their fundamental blocks, which could take the form of a parallepiped or a hexagonal prism for a Euclidian space or more complicated polyhedrons for spherical and hyperbolic spaces. The way the faces of these blocks are glued together also differs between each space. The surprise from the WMAP data is that the topology of space seems indeed to be multiply connected, and described by a special class of shapes that are called "well proportioned".

> **At a glance : Cosmic topology**
>
> • There are three possibilities for the curvature of the universe: space can be flat, spherical or hyperbolic.
> • The geometry of the universe depends on its curvature and also on its topology, which governs the way space is connected and so determines if it is finite or infinite
> • Measurements of the cosmic microwave background constrain the curvature of the universe and provide hints about its topology
> • Recent data suggest that the universe might be multiply connected, like the left- and right-hand sides of the screen in a computer game
> • Since the early 1990s the number of cosmo-topologists around the world has grown to more than 50

**Cosmic harmonics**

The best way to determine the shape of our universe is to go back to its beginning, just after the Big Bang. The infant universe is thought to have been crossed by real acoustic waves that would have caused tiny density fluctuations in the primordial plasma. After about 380::000 years, however, the universe had expanded and cooled enough to allow hydrogen and helium to form. This meant that photons could suddenly travel unhindered through space, carrying with them vital information about the primordial density fluctuations (which are now thought to have been the seeds for galaxies and clusters of galaxies to form). Today, 13.7 billion years after the Big Bang, this radiation has cooled to a temperature of about 2.7 K, which is in the microwave region. And the fluctuations are imprinted as hot and cold spots in this cosmic microwave background.

A good way to understand the connection between acoustics and topology is to sprinkle fine sand uniformly over a drumhead and then make it vibrate. The grains of sand will collect in characteristic spots and patterns that reveal information about the local geometry of the drum and about the elasticity of its membrane. But the distribution of spots also depends on the global shape -- i.e. the topology -- of the drum. For example, the waves will be reflected differently according to whether the drumhead is infinite or finite, and whether it is shaped like a circle, an ellipse or some other shape.

Just as the vibration of a drumhead may be expressed as a combination of its harmonics, fluctuations in the temperature of the cosmic background radiation may be expressed as combinations of the vibrational modes of space itself. When the level of fluctuations is plotted as a function of angle we therefore find a series of peaks that provide a signature of the geometry of space 13.7 billion years ago (figure 2). For example, the position and amplitude of the first peak -- i.e. the peak at the largest angle -- in this "angular power spectrum" gives the radius of curvature of space.

Different cosmological models predict different power spectra, and high-resolution measurements of the cosmic microwave background from instruments such as WMAP now allow us to compare different theories against real data. However, when WMAP released its first data in 2003, tenants of the standard cosmological model were faced with several surprises.

The position of peaks in the angular spectrum is usually described by their wavenumber or mode $l = 180°/\theta$, where $\theta$ is the angular distance in the sky. In fact, the lowest $l = 1$ or "dipole" mode is swamped by the far stronger dipole induced by the motion of the solar system relative to the cosmic background, which means that it cannot be measured. But when researchers determined the first observable mode -- the $l = 2$ "quadrupole" -- they found that it was seven times weaker than expected for a flat, infinite universe. Furthermore, the "octopole mode" with $l = 3$ was also found to be weaker than the expected value, by a factor of about two thirds.

For higher modes up to $l = 900$, which correspond to angular scales of just 0.2°, the WMAP data were fairly consistent with the standard model. But a more careful analysis of the power spectrum also revealed that the distribution of temperature fluctuations is not fully isotropic and that the fluctuations are distributed differently on different angular sales.

All these anomalies contradict the standard picture of the universe, which has led some more conservative cosmologists to claim that they are due to bad data analysis. Furthermore, the second round of WMAP data -- originally expected in February 2004 -- has been delayed for more than a year, which may hint at additional trouble to come! Meanwhile, other cosmologists have taken the problem seriously and proposed new laws to explain the early universe, some of which have exotic names such as "vanilla" and "racetrack" inflation.

Cosmo-topologists, on the other hand, have tried to find a more natural geometrical explanation for the observed power spectrum. Put simply, the unusually low amplitudes of the quadrupole and octopole modes means that long wavelengths (i.e. temperature fluctuations over large angular scales) are missing, which could simply be because space is not big enough to sustain them. This can be likened to oscillation of a string fixed at both ends, where the maximum wavelength of an oscillation is twice the string length. The geometrical explanation of the power spectrum thus implies that we line in a finite, multiply-connected space that is smaller than the observable universe.

**Dodecahedral space**

Surprisingly, not all small-volume universes suppress the large-scale fluctuations. In 2003 the present author, Jeff Weeks and co-workers proved that the long-wavelength modes tend to be relatively lowered only in a special family of finite, multiconnected spaces that are called "well-proportioned spaces" because they have a similar extent in all three dimensions. More specifically, we discovered that the best candidate to fit the observed power spectrum is a well-proportioned space called the Poincaré dodecahedral space.

This space may be represented by a polyhedron with 12 pentagonal faces, with opposite faces being "glued" together after a twist of 36° (figure 3). This is the only consistent way to obtain a spherical (i.e. positively curved) space from a dodecahedron: if the twist was 108°, for example, we would end up with a radically different hyperbolic space. The Poincaré dodecahedral space is essentially a multiply connected variant of a simply connected hypersphere, although its volume is 120 times smaller.

A rocket leaving the dodecahedron through a given face immediately re-enters through the opposite face, and light propagates such that any observer whose line-of-sight intercepts one face has the illusion of seeing a slightly rotated copy of their own dodecahedron. This means that some photons from the cosmic microwave background, for example, would appear twice in the sky.

The power spectrum associated with the Poincaré dodecahedral space is different from that of a flat space because the fluctuations in the cosmic microwave background will change as a function of their wavelengths. In other words, due to a cut-off in space corresponding to the size of the dodecahedron, one expects fewer fluctuations at large angular scales than in an infinite flat space, but at small angular scales one must recover the same pattern as in the flat infinite space. In order to calculate the power spectrum we varied the mass–energy density of the dodecahedral universe and computed the quadrupole and the octopole modes relative to the WMAP data. To our delight, we found a small interval of values over which both these modes matched the observations perfectly. Moreover, the best fit occurred in the range $1.01 < \Omega < 1.02$, which sits comfortably with the observed value.

The Poincaré dodecahedral space therefore accounts for the lack of large-scale fluctuations in the microwave background and also for the slight positive curvature of space inferred from WMAP and other observations. Moreover, given the observed values of the mass–energy densities and of the expansion rate of the universe, the size of the dodecahedral universe can be calculated. We found that the smallest dimension of the Poincaré dodecahedron space is 43 billion lightyears, compared with 53 billion light-years for the "horizon radius" of the observable universe. Moreover, the volume of this universe is about 20% smaller than the volume of the observable universe. (There is a common misconception that the horizon radius of a flat universe is 13.7 billion light-years, since that is the age of the universe multiplied by the speed of light. However, the horizon radius is actually much larger because photons from the horizon that are reaching us now have had to cross a much larger distance due to the expansion of the universe.)

If physical space is indeed smaller than the observable universe, some points on the map of the cosmic microwave background will have several copies. As first shown by Neil Cornish of Montana State University and co-workers in 1998, these ghost images would appear as pairs of so-called matched circles in the cosmic microwave background where the temperature fluctuations should be the same (figure 4). This "lensing" effect, which can be precisely calculated, is thus purely attributable to the topology of the universe.

Due to its 12-sided regular shape, the Poincaré dodecahedral model actually predicts six pairs of diametrically opposite matched circles with an angular radius of 10–50°, depending on the precise values of cosmological parameters such as the mass–energy density.

**Circles in the sky**

When news of our dodecahedral model appeared in *Nature* in October 2003, it was not long before the press started running headlines based on what was being hailed the new "football-shaped" model of the universe. However, since cosmo-topology is a very competitive
field, the initial response from other groups was not always favourable.

For instance, the *New York Times* ran the headline "Cosmic soccer ball? Theory already takes sharp kicks", based on an apparently negative search for matched circles in the WMAP data performed by Cornish and coworkers. Using massive computer simulations, they claimed to have found no evidence of matching on angular sizes greater than 25° and thus rejected the Poincaré hypothesis the same day it appeared.

In fact, their rejection was rather premature because they had only looked for non-rotated matched circles that were diametrically opposite one another – a case that did not test the dodecahedron model at all. After the initial excitement, Cornish and co-workers went back and reassessed the data. Taking account of the additional 36° twist took a few additional months of computer time, but the matched circles remained elusive. This led them to conclude that there was no reasonable topology for the universe that had a characteristic length smaller than the observable horizon.
However, it turned out that the researchers had taken a short cut to save computer time. While they correctly took into account the possible rotations between matched circles that are implied by most multi-connected topologies, they only searched for matched circles that were back-to-back or very nearly back-to-back. This led them to exclude *all* likely multiply connected spaces. In the mean time, however, we had proved that in most multiply connected, well-proportioned topologies space is not homogeneous. This means that the position of matched circles in the sky depends on the location of the observer, and they are not, therefore, back-to-back. Only in the simplest of topologies, such as the hypertorus in flat space and the Poincaré dodecahedron in a spherical space, is space homogeneous and the circles back-to-back.

This violates one of the most basic principles of cosmology, that there is no privileged position in the universe. But this principle could be illusory, like the ant in the desert that is convinced the whole world is filled with sand and dunes. For instance, in a flat-torus universe, any gluing together of the opposite faces combined with a screw motion produces pair of circles that are far from being back-to-back. Unfortunately, the increase in the number of degrees of freedom that results from such a scenario means that a full-circle search in the WMAP data is beyond current computing capabilities.

**Cosmic horn**

In June 2004, however, Boud Roukema and colleagues at the Torun Centre for Astronomy in Poland independently searched for circles in the WMAP data. By only looking for back-to-back circles within a limited range of angular sizes and neglecting all other possible matches, the computer time was reduced drastically. Remarkably, the Polish team found six pairs of matched circles distributed in a dodecahedral pattern and twisted by 36°, each with an angular size of about 11°. This implied that $\Omega = 1.010 \pm 0.001$, which is perfectly consistent with our dodecahedral model, although the result was much less publicized than the earlier negative results.

In fact, the statistical significance of the match still needs to be improved, which means that the validity of the Poincaré dodecahedron model is still open to debate. In the last few months, however, there has been much theoretical progress on well-proportioned spaces in general. Early this year, for example, Frank Steiner and co-workers at the University of Ulm in Germany proposed a multiply connected hyperbolic topology called the Picard hyperbolic space. Like the Poincaré dodecahedron, this horn-shaped space belongs to the family of well-proportioned spaces and it also correctly fits the low vibrational modes of the WMAP data. However, since the topology requires the density parameter to have a value of $\Omega = 0.95$, and thus a negatively curved space, it does not fit the experimental constraints we already have on the curvature of space.

After studying the horn-shaped topology further, Steiner and co-workers realized that well-proportioned spherical spaces were, in fact, more promising. They went on to prove that the fit between the power spectrum predicted by the Poincaré dodecahedron model and that observed by WMAP was even better than we had previously thought. But the German team also extended its calculations to well-proportioned tetrahedral and octahedral spherical spaces in which $\Omega > 1$ (see figure 3).

These spaces are somewhat easier to understand than a dodecahedral space, but they require higher values of the density: $\Omega > 1.015$ for octahedral spaces and $\Omega > 1.025$ for tetrahedral spaces, compared with $\Omega > 1.009$ for dodecahedral spaces. However, these values are still compatible with the WMAP data. Furthermore, Steiner and co-workers found that the signal for pairs of matched circles could have be missed by current analyses of the cosmic microwave background due to various measurement effects that damage or even destroy the temperature matching.

Another active area of cosmic topology is "cosmic crystallography", which was initially devised by the present author and co-workers in 1996 and is now being pursued by, among others, Germán Gomero of the Universidade Estadual Paulista in Brazil and Marcelo Reboucas of the Brazilian Center for Research in Physics. In cosmic crystallography researchers look for repeating patterns in the 3D distribution of high-redshift sources, such as galaxy clusters and quasars, much like the repeating patterns of atoms observed in crystals. By building so-called pair-separation histograms, cosmologists are in most cases able to detect a multiconnected topology of space in the form of spikes that clearly stand out above the distribution expected for the simply connected case.

**A Pandora's box for physics**

Finite well-proportioned spaces, especially the Poincaré dodecahedron, open something of a Pandora's box for the physics of the early universe. The standard model of cosmology relies in the main on the hypothesis that the early universe underwent a phase of exponential expansion called inflation, which produced density fluctuations on all scales. In the simplest inflationary models, space is supposed to have become immensely larger than the observable universe. Therefore, a positive curvature (i.e. $\Omega > 1$), even if weak, implies a finite space and sets strong constraints on inflationary models.

It is possible to build "low scale" inflationary universes in which the inflation phase ends more quickly than it does in general inflationary modes, leading to a detectable space curvature. In other words, even if space is not flat, a multiconnected topology does not contradict the general idea of inflation. However, no convincing physical scenario for this has yet been proposed.

Perhaps the most fundamental challenge is to link the present-day topology of space to a quantum origin, since general relativity does not allow for topological changes during the course of cosmic evolution. A quantum theory of gravity could allow us to address this problem, but there is currently no indication about how such a unified theory might actually describe the emergence of multiply connected spaces.

Data from the European Planck Surveyor, which is scheduled for launch in 2007, will be able to determine $\Omega$ with a precision of 1%. A value lower than 1.01 will rule out the Poincaré dodecahedron model, since the size of the corresponding dodecahedron would become greater than the observable universe and would not leave any observable imprint on the microwave background. A value greater than 1.01, on the other hand, would strengthen the models' cosmological pertinence.

Whether or not some multiply connected model of space such as the Poincaré dodecahedron is refuted by future astronomical data, cosmic topology will continue to remain at the heart of our understanding about the ultimate structure of our universe.

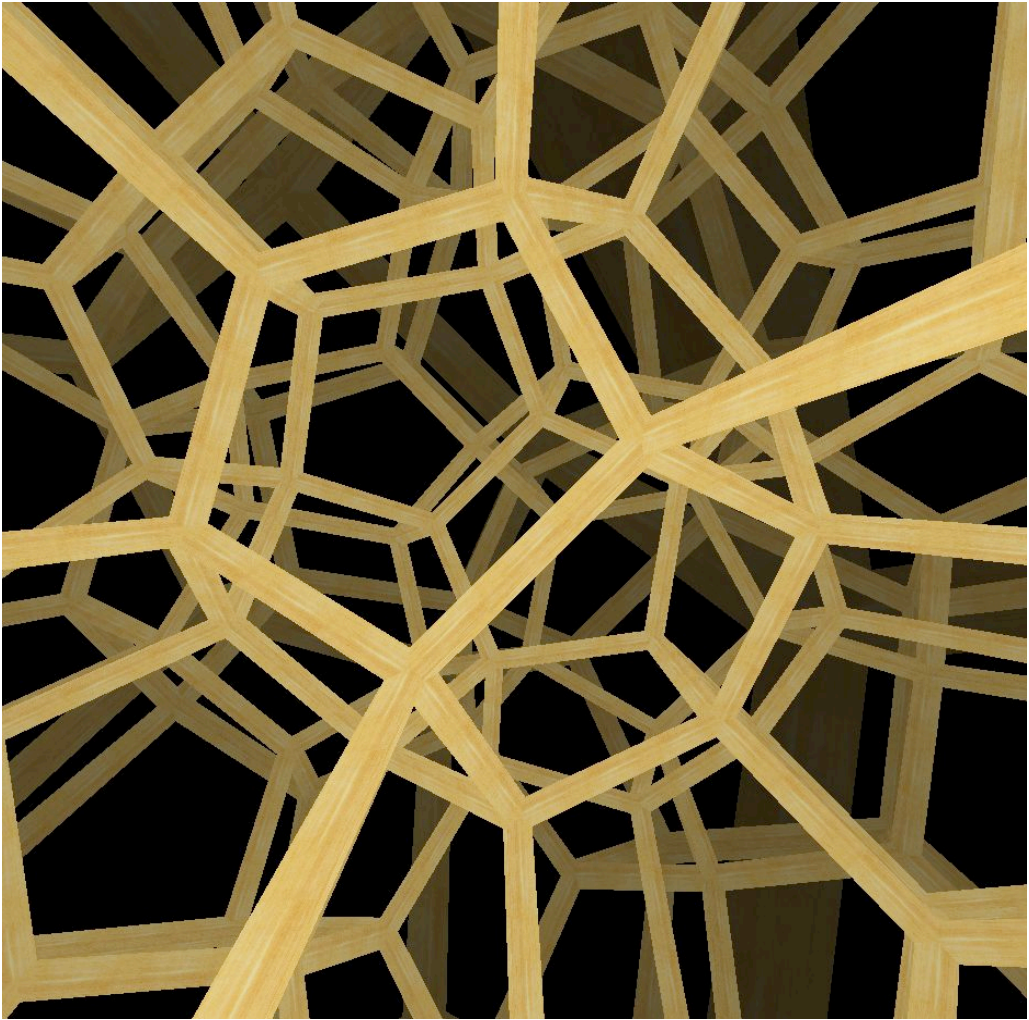

**Lost in space**
A multiconnected dodecahedral universe gives the impression of being 120 times bigger than it actually is, similar to a hall of mirrors [Credit: Jeff Weeks].

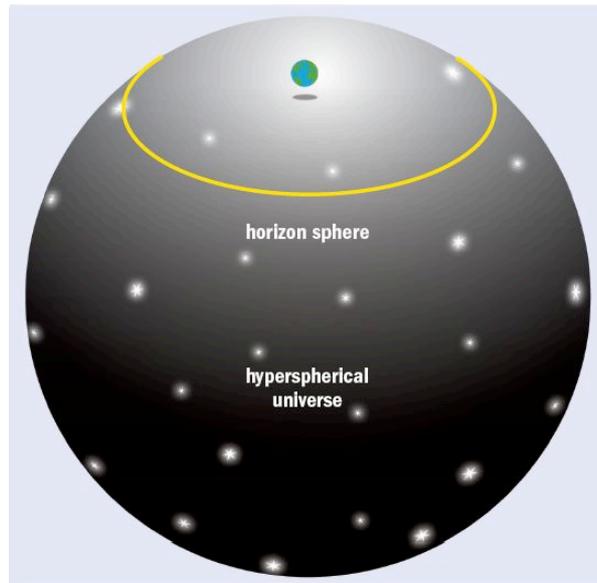

**Figure 1: How flat is the universe?**

The curvature of space and our horizon radius is determined by the average density of the universe and its expansion rate. Cosmologists often say that space is *nearly* flat because the observed value of the density is close to the critical value for a flat universe. However, if the density was 2% more than the critical density, the horizon radius (red line) would be about 46 billion light-years and the radius of curvature of the corresponding hypersphere would be only 2.6 times greater. We would therefore see a modest but non-trivial portion of the hypersphere. If the density is exactly equal to the critical value, space is Euclidean, the radius of curvature is infinite and we can only see an infinitesimal portion of the universe.

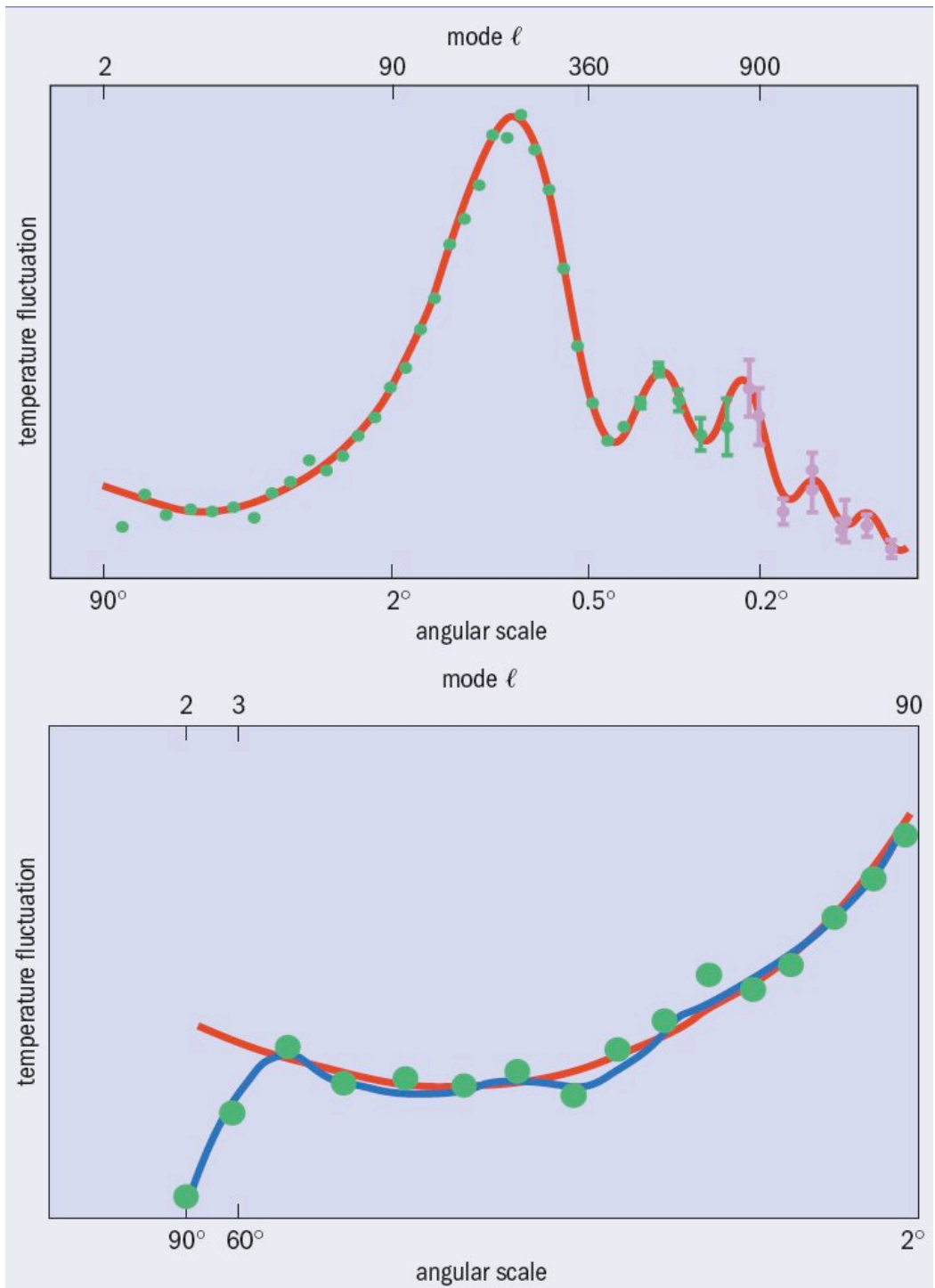

**Figure 2: Angular power spectrum**
The evidence that space might have an unusual shape can be seen when we plot the level of temperature fluctuations in the cosmic microwave background as a function of angle, θ (the peaks in the spectrum corresponds to modes l = 180°/θ). In particular, the largest angular scales (or lowest modes) are very sensitive to the curvature of space (inset). In this region data from WMAP (black dots) do not agree with theoretical predictions for an infinite Euclidean space (red curve), whereas predictions based on a Poincaré dodecahedral space (blue curve) do agree. This suggests that we live in a multiply-connected spherical space.

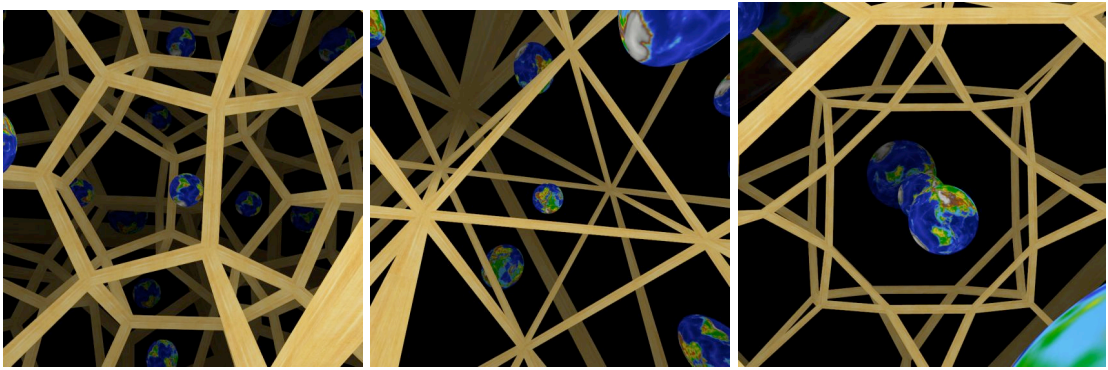

**Figure 3: The many shapes of the universe**

The Poincaré dodecahedral space (left) can be described as the interior of a "sphere" made from 12 slightly curved pentagons. However, this shape has a big difference compared with a football because when one goes out from a pentagonal face, one comes back immediately inside the ball from the opposite face after a 36° rotation. Such a multiply connected space can therefore generate multiple images of the same object, such as a planet or a photon. Other such spaces that fit the WMAP data are the tetrahedron (middle) and octahedron (right). [Credit: Jeff Weeks]

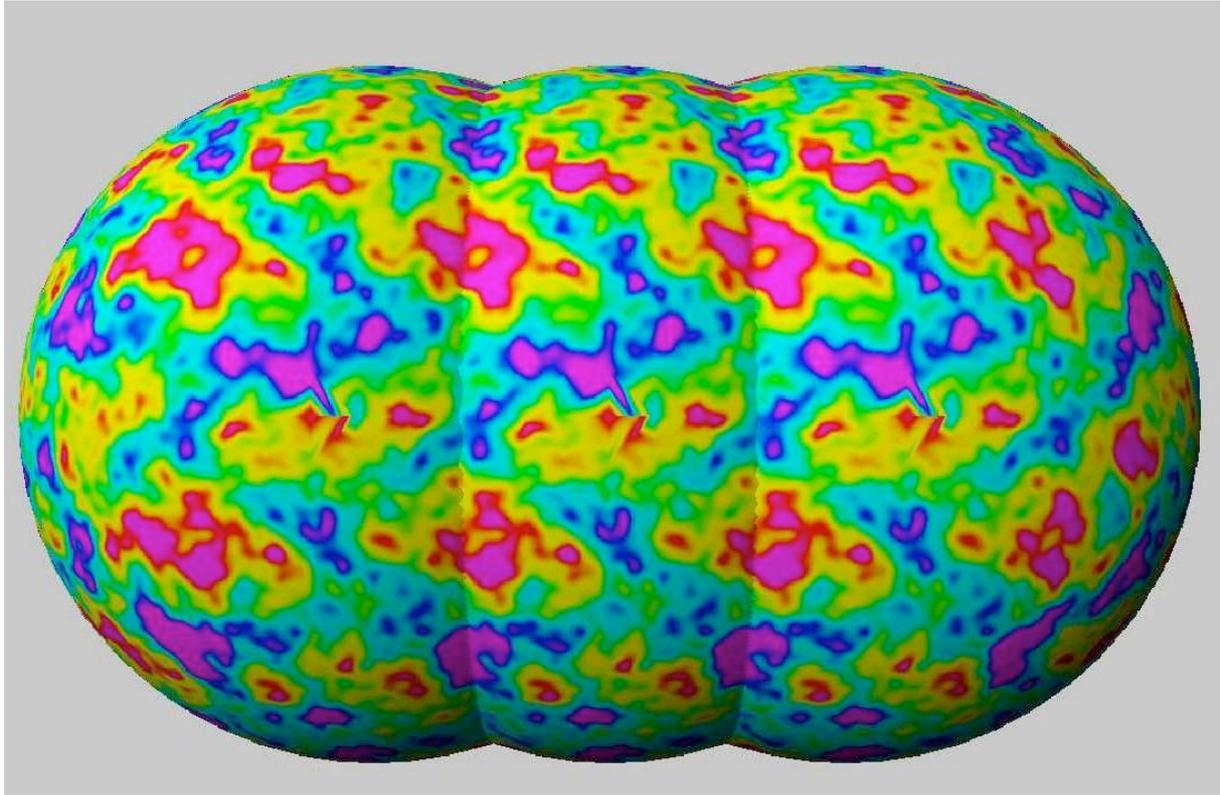

**Figure 4: Simulated circle matching**

The topology of the universe describes how different regions are connected and should therefore leave its imprint on the cosmic microwave background. For example, if our physical space is smaller than the observable universe (as recent data suggest it is) then the horizon sphere wraps around the universe and intersects itself. As a result, duplicated images of the cosmic microwave background (in which the colours represent temperature fluctuations) will intersect along a circle and we would observe this circle on different sides of the sky. [Credit: A Riazuelo]